\documentclass[aps,prl,reprint,groupedaddress]{revtex4-1}

\usepackage{graphicx,amsmath,color,amsfonts,bm}
\newcommand{\be}{\begin{equation}}
\newcommand{\ee}{\end{equation}}

\begin{document}

\title{Berry phase jumps and giant nonreciprocity in Dirac quantum dots}

\author{Joaquin F. Rodriguez-Nieva and Leonid S. Levitov}
\affiliation{Department of Physics, Massachusetts Institute of Technology, Cambridge, MA 02139, USA}

\begin{abstract}
We predict that a strong nonreciprocity in the resonance spectra of Dirac quantum dots can be induced by the Berry phase. The nonreciprocity arises in relatively weak magnetic fields  and is manifest in anomalously large field-induced splittings of quantum dot resonances which are degenerate at $B=0$ due to time-reversal symmetry. This exotic behavior, which is governed by field-induced jumps in the Berry phase of confined electronic states, is unique to quantum dots in Dirac materials and is absent in conventional quantum dots. The effect is strong for gapless Dirac particles and can overwhelm the $B$-induced orbital and Zeeman splittings. A finite Dirac mass suppresses the effect. The nonreciprocity, predicted for generic two-dimensional Dirac materials, is accessible through Faraday and Kerr optical rotation measurements and scanning tunneling spectroscopy. 
\end{abstract}



\maketitle

\section{I. Introduction}

Quantum dots can be embedded in two-dimensional Dirac systems using local gate potentials and point charges, as recently demonstrated in graphene\cite{zhaoscience,crommie}. These Dirac quantum dots are defined by nanoscale p-n-junction rings, with Klein scattering at the p-n junctions serving as a vehicle for confinement of electronic states\cite{fockstates,causticsqd,integrableqd,quantumdot,qdboundstates,scatteringqd}. Carrier confinement in these ring-shaped electron resonators arises due to constructive interference of electronic waves scattered at the p-n junction\cite{klein1,klein2} and inward-reflected from the ring. Confined states are manifest through resonances appearing periodically in scanning tunneling spectroscopy maps\cite{zhaoscience,crommie}. 

Here we show that this mechanism for electronic confinement can be exploited for accessing exotic and potentially useful behavior which is not available in conventional quantum dots. In particular, we predict that the Berry phase, a signature topological characteristic of Dirac materials\cite{berry,changbp,graphene4, graphene3, changreview,niu2,niu}, induces strong nonreciprocity of quantum dot resonances in the presence of a weak magnetic field $B$:
\be
\varepsilon_{n,m} \neq \varepsilon_{n,-m}.
\ee
Here $m$ and $n$ denote the azimuthal and radial quantum numbers, respectively (for optical nonreciprocity, see Refs.\cite{fan,ross}). As we will see, resonance splittings of the $\pm m$ states, which are degenerate at $B=0$, grow rapidly with magnetic field, approaching values as large as half the quantum dot resonance period $\Delta\varepsilon$. In particular, for the weak $B$ of interest, the effect dominates over the $B$-induced orbital and Zeeman splittings. 

\begin{figure}[b]
\centering \includegraphics[scale=1.0]{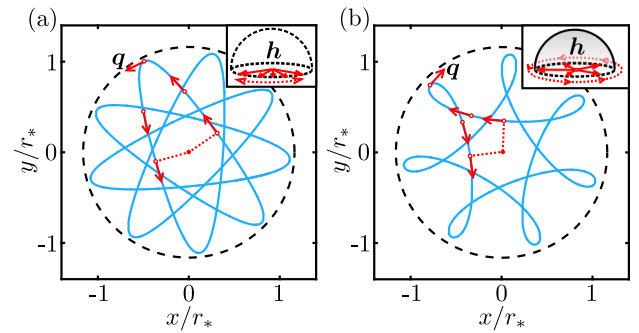}
\vspace{-7mm}
\caption{Controlling the Berry phase of confined Dirac electrons using magnetic fields. Shown are semiclassical orbits of a massless particle exhibiting topologically distinct orbital behavior corresponding to (a) $B < B_{\rm c}$ and (b) $B > B_{\rm c}$ [see critical field $B_{\rm c}$ in Eq.(\ref{eq:bstar})]. The Berry phase, determined by the solid angle subtended by ${\bm h}=(h_x,h_y,h_z)$ in Eq.(\ref{eq:bp}), jumps from $\varphi_{\rm B} = 0$ to $\varphi_{\rm B} = \pi$ at $B=B_{\rm c}$, see insets [for gapless systems ${h}_{x,y} = v\,q_{x,y}$ and $h_z = 0$, with ${q}_{x,y}$ the kinetic momentum (red vectors) and $v$ the Fermi velocity]. Here we used $m = 1 / 2$, energy $\varepsilon = 1.35\, \hbar v/r_*$, with $r_*$ defined in Eq.(\ref{eq:units}), $B / B_{\rm c} = 0.8$ for (a) and  $B/B_{\rm c}= 1.6 $ for (b).}
\label{fig:schematics}
\end{figure}

\section{II. Semiclassical Description}

The Berry phase jumps can be understood from a simple semiclassical picture describing confined electrons in a gapless two-band system. For a confining potential with circular symmetry, the resonance spectrum $\varepsilon_{n,m}$ can be obtained from the WKB condition for $\varphi_{\rm orb} = \frac{1}{\hbar}\oint_{{\cal C}} d{\bm r}\cdot {\bm p}$, the usual orbital phase accumulated along the classical path ${\cal C}$: 
\be
\varphi_{\rm orb}(\varepsilon,m) + \varphi_{\rm B}(\varepsilon,m) = 2\pi (n+\gamma),
\label{eq:wkb}
\ee
where $\varphi_{\rm B}$ is Berry phase and $\gamma$ is a constant\cite{niu2,niu,stone}. The Dirac band structure, viewed as a Zeeman-type Hamiltonian ${\cal H} = {\bm h}({\bm p})\cdot{\boldsymbol\sigma}$, where ${\boldsymbol\sigma}=(\sigma_x,\sigma_y,\sigma_z)$ are Pauli matrices, gives rise to a geometric gauge field that generates the Berry phase,
\be
\varphi_{\rm B} = \oint_{\cal C} d{\bm p} \cdot \langle {\bm h}_+ | i \nabla_{\bm p} |{\bm h}_+ \rangle = S({\cal C})/2.
\label{eq:bp}
\ee
In Eq.(\ref{eq:bp}), $S({\cal C})$ denotes the solid angle subtended by the vector ${\bm h} = (h_x,h_y,h_z)$ along a closed path ${\cal C}$, and $|{\bm h}_\pm\rangle$ are eigenstates of the two-band Hamiltonian:
\be
{\cal H}|{\bm h}_\pm\rangle = \pm |{\bm h}| |{\bm h}_\pm\rangle.
   \label{eq:hparameters}
\ee
The Berry phase in a gapless system ($h_z=0$) can only take the values $\varphi_{\rm B} = 0$ or $\pm\pi$ \cite{graphene1,graphene2,graphene3,graphene4}. 

\begin{figure}[t]
  \centering \includegraphics[scale=1.0]{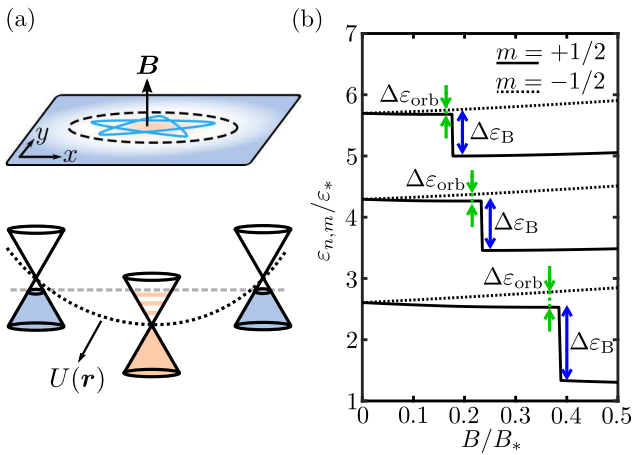}
  \vspace{-7mm}
  \caption{Magnetic response of quantum dot resonances in a gapless Dirac system. (a) The quantum dot is defined by the circular p-n ring (dashed lines) induced by a radial electrostatic potential $U({\bm r})$. (b) The magnetic response is dominated by the Berry-phase splitting $\Delta\varepsilon_{\rm B}$, which is approximately half the resonance period $\Delta\varepsilon$. Also indicated in the figure is the orbital splitting $\Delta\varepsilon_{\rm orb}$. Peak splitting is calculated from Eq.(\ref{eq:wkb}) for $n=0,1,2$, $m = \pm 1/2$, and $\gamma = 0.6$; $B_{\rm c}$ is calculated from Eq.(\ref{eq:bstar}); $\varepsilon_*$ and $B_*$ are defined in Eq.(\ref{eq:units}). 
}
\label{fig:splitting}
\vspace{-5mm}
\end{figure}

An external magnetic field can alter the Berry phase of the orbits, allowing them to switch between the $\varphi_{\rm B} = 0$ and $\pm\pi$ types. As illustrated in Fig.\ref{fig:schematics}, switching can take place even in a weak magnetic field. In particular, for $B=0$ we find $\varphi_{\rm B}(\varepsilon,\pm m)=0$, whereas for weak nonzero fields we find $\varphi_{\rm B}(\varepsilon,m)=\pi$ and $\varphi_{\rm B}(\varepsilon,-m)=0$. As a result of the $\pi$ difference in the WKB condition in Eq.(\ref{eq:wkb}) for the $\pm m$ states, the $m>0$ and $m<0$ families of resonances are shifted by half a period, giving rise to a large resonance splitting (Fig.\ref{fig:splitting}): 
\be
\varepsilon_{n,m} - \varepsilon_{n,-m} \approx \Delta\varepsilon / 2, \quad ({\rm gapless})
\label{eq:gapless}
\ee
where $\Delta\varepsilon \sim 10{\rm -}50\,{\rm meV}$ is the spacing of resonances in each family. Equation (\ref{eq:gapless}) describes gapless Dirac bandstructures, a generalization for gapped systems is discussed below. 

\begin{figure}
\centering \includegraphics[scale=1.0]{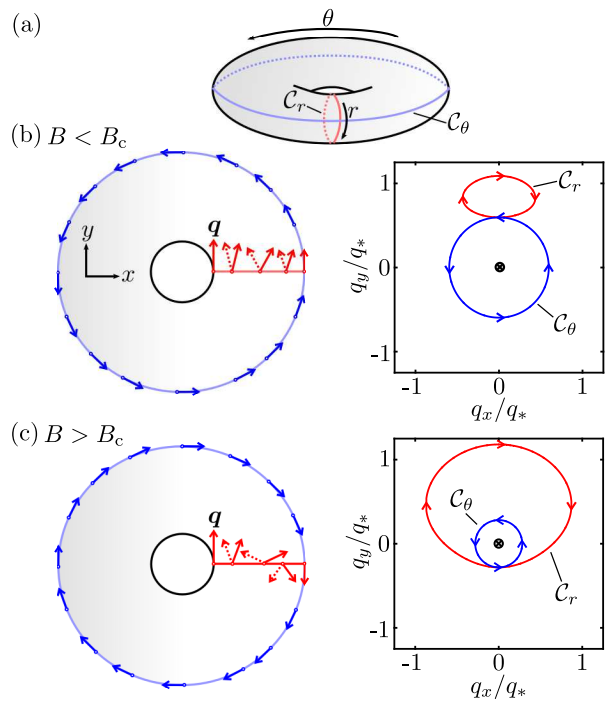}
\caption{Topologically distinct mappings of ${\bm q}$ [Eq.(\ref{eq:qr})] to the surface of a torus (a), plotted for (b) $B<B_{\rm c}$ and (c) $B>B_{\rm c}$. Indicated with blue(red) arrows is ${\bm q}$ along the curves ${\cal C}_\theta$(${\cal C}_r$) shown in panel (a), where dotted lines/arrows indicate a curve/vector in the bottom surface of the torus. At $B = B_{\rm c}$, there is a transition between trivial and non-trivial winding of ${\bm q}$ along ${\cal C}_r$. This results in $B$-induced phase jumps of the Berry phase. Here we define $q_* = \varepsilon_*/ v$ and use the same parameter values as in Fig.\ref{fig:schematics}.}
\label{fig:ebk}
\vspace{-5mm}
\end{figure}

To illustrate how $B$ controls the Berry phase, we consider a massless particle confined in a radial electrostatic potential $U(r)$. This corresponds to ${\bm h} = v (q_x,q_y,0)$ in Eq.(\ref{eq:hparameters}). In the presence of a uniform magnetic field $B$, the kinetic momentum ${\bm q} = {\bm p} - e{\bm A}$ is given by
\be
\begin{array}{l}
\displaystyle q_r = p_r = \pm\sqrt{\left[\varepsilon - U(r)\right]^2/v^2-\left(m\hbar/r - eBr\right/2)^2}, \\
\displaystyle q_\theta = p_\theta - eA_\theta = m\hbar / r - eBr/2. 
\end{array}
\label{eq:qr}
\ee
Here $v$ is the electron velocity, and we used the axial gauge $A_x = -By/2$, $A_y = Bx/2$ to preserve rotational symmetry. Because the system is integrable, with constants of motion $\varepsilon$ and $m$, we can map ${\bm q}$ to the surface of a torus. Figure \ref{fig:ebk} shows such mapping, with ${\bm q}$ plotted along two curves: ${\cal C}_\theta$ in the toroidal direction and ${\cal C}_r$ in the poloidal direction. At a critical $B = B_{\rm c}$ we find that the winding number of ${\bm q}$ along ${\cal C}_r$ jumps from 0 to 1, thus resulting in a $\pi$-jump of ${\varphi}_{\rm B}$.

The semiclassical quantization of quantum dot resonances can now be obtained from Eq.(\ref{eq:wkb}) using ${\bm q}$ in Eq.(\ref{eq:qr}) evaluated on both ${\cal C} = {\cal C}_{\theta}$ and ${\cal C} = {\cal C}_{r}$ \cite{bs}. This yields two quantization conditions for $m$ and $\varepsilon$. For ${\cal C} = {\cal C}_\theta$, Eq.(\ref{eq:wkb}) yields $m = n_{\theta}+\gamma_\theta - \varphi_{\rm B}/ 2 \pi$, where $\varphi_{\rm B} = \pi$ independently of $B$ [see blue curves in panels (b) and (c) of Fig.\ref{fig:ebk}]. Using $\gamma_\theta = 0$, we find the anticipated quantization of angular momentum $m$=half--integer. For ${\cal C} = {\cal C}_r$, instead, we find $\frac{1}{\hbar}\int_{r_1}^{r_2}dr\,p_r = 2 \pi (n_r + \gamma_r) -\varphi_{\rm B}$, where $r_1$ and $r_2$ are the classical return points. The half period shift in the radial quantization condition results from the $\pi$-jump in $\varphi_{\rm B} $ at $B=B_{\rm c}$.

While the same semiclassical picture applies to gapped Dirac systems ($h_z \ne 0$), there are important differences with respect to the gapless case. In particular, the solid angle subtended by the vector ${\bm h}({\bm q})$, which now points towards the upper hemisphere, is strictly smaller than $2\pi$; nonreciprocity induced by Berry phase is quenched at increasing bandgaps, as will be shown with a more detailed quantum model in Fig.\ref{fig:gapeffect}. In the limit $|h_z| \gg |h_{x,y}|$, orbital splitting dominates. 

The jump in Berry phase corresponds to a transition from convex orbits to skipping orbits (Fig.\ref{fig:schematics}). This observation allows to define the critical field $B_{\rm c}$ that induces giant nonreciprocity, i.e. the field necessary to reverse the electron velocity at the outer classical return point. From Eq.(\ref{eq:qr}) we find $ q_\theta = m\hbar / r_2(\varepsilon) - e B_{\rm c} r_2 (\varepsilon) /2  =0 $, with $r_2(\varepsilon) $ the outer return point [i.e., $q_r (r_2) = 0$]. For a quadratic potential model $U({r}) = \kappa {r}^2$, this condition yields 
\be
B_{\rm c}[{\rm T}] = \frac{2 \hbar m \kappa}{e\varepsilon} = 1.3\, \frac{m \kappa [{\rm eV}/\mu{\rm m^2}]}{\varepsilon[{\rm meV}]}.
\label{eq:bstar}
\ee
Using typical values corresponding to recent experiments\cite{zhaoscience}, $\kappa \approx 4\,{\rm eV}/\mu{\rm m^2}$, $\varepsilon \approx 10\,{\rm meV}$ and $m=1/2$, we find $B_{\rm c}$ on the order of $0.3\,{\rm T}$.

Besides the splitting arising at $B=B_{\rm c}$, another signature of the nonreciprocal effect is the linear $m$-dependence of $B_{\rm c}$, see Eq.(\ref{eq:bstar}). This dependence can be understood by noticing that, for larger $m$, a larger $B$ is necessary to induce skipping orbits. As we will see, the $m$ dependence of $B_{\rm c}$ gives rise to a peculiar branching pattern of the quantum dot resonances which can be probed in spectral measurements away from the quantum dot center (see Sec.III).

Importantly, the giant nonreciprocal effect relies on the splitting due to Berry phase being dominant over orbital and Zeeman splittings. This is the case, in particular, for the value $B_{\rm c} \sim 0.3\,{\rm T}$ found in Eq.(\ref{eq:bstar}). Indeed, $B_{\rm c}$ is significantly lower than the value $B_{\rm LL} = (\Delta\varepsilon)^2 / e \hbar v_{\rm F}^2 \sim 1\,{\rm T}$ which is necessary for the first Landau level to be larger than the resonance period $\Delta \varepsilon \approx 25\,{\rm meV}$. The strength of the nonreciprocal effect is illustrated in Fig.\ref{fig:splitting} which shows the semiclassical spectrum obtained from Eq.(\ref{eq:wkb}) for $n=0,1,2$ and $m=\pm 1/2$ including both orbital and Berry phase splitting. For typical model parameters, the splitting $\Delta\varepsilon_{\rm B}\sim\Delta\varepsilon/2$ induced by the Berry phase jump dominates over the conventional orbital splitting $\Delta\varepsilon_{\rm orb}$. The effect becomes more dramatic at larger $n$ and smaller $m$. Furthermore, the energy $\varepsilon_{\rm Z}$ for the electron Zeeman splitting, $\varepsilon_{\rm Z} = \mu_{\rm B}B_{\rm c} \sim 10^{-2}\,{\rm meV}$, is negligible compared to the characteristic energy of quantum dots (here $\mu_{\rm B} \approx 5.8\cdot 10^{-5}\,{\rm eV/T} $ is the Bohr magneton).

\section{III. Microscopic Model}

To supplement the simple semiclassical picture above with a microscopic quantum model, we consider the Dirac equation describing confined electrons in the presence of a uniform magnetic field:
\be
\left[ v\,{\boldsymbol\sigma}\cdot {\bm q} + (\Delta/2) \sigma_z+U({\bm r}) \right] \Psi({\bm r}) = \varepsilon \Psi({\bm r}).
\label{eq:H}
\ee
Here $\Delta$ is the bandgap and ${\bm q}$ is the kinematic momentum with components $q_{x,y} = -i \hbar \partial_{x,y} -eA_{x,y}$ and $q_z =0 $. This corresponds to ${\bm h} = v (q_x,q_y,\Delta/2v)$ in Eq.(\ref{eq:hparameters}). Because we are interested in eigenstates confined inside the p-n ring, with radius smaller than the characteristic length of the electrostatic potential, it is legitimate to use a parabolic potential model $U(\bm r)\approx\kappa {\bm r}^2$. By using the axial gauge $A_x = -By/2$, $A_y = Bx/2$ to preserve rotational symmetry, the eigenstates of Eq.(\ref{eq:H}) can be expressed using the polar decomposition ansatz, 
\be
\Psi_m (r,\theta) = \frac{e^{im\theta}}{\sqrt{r}} \left( \begin{array}{c} u_{1}(r) e^{-i\theta/2} \\ i u_{2}(r) e^{i\theta/2} \end{array}\right),
\label{eq:harmonic}
\ee
with $m$ a half-integer number. This decomposition allows to rewrite Eq.(\ref{eq:H}) as 
\be
\left( \begin{array}{cc} r^2 - \varepsilon + \Delta / 2  & \partial_{r} + m/r - B r/2 \\ -\partial_{r}  + m/r - Br/2 & r^2 - \varepsilon -\Delta/2  \end{array} \right) \left(\begin{array}{c} u_1 \\ u_2 \end{array}\right) = 0.
\label{eq:Hnd}
\ee
Here $r$ and $B$ are in units of $r_*$ and $B_*$, respectively, whereas $\varepsilon$ and $\Delta$ are in units of $\varepsilon_*$, with 
\be
\begin{array}{l}
r_* = \sqrt[3]{\hbar v/\kappa} \sim 60\,{\rm nm},\quad
\varepsilon_* = \sqrt[3]{(\hbar v)^2 \kappa} \sim 10\,{\rm meV}, \\ \\
B_* = (\hbar / e)\cdot \sqrt[3]{(\kappa/\hbar v)^2} \sim 0.2\,{\rm T}.
\end{array}
\label{eq:units}
\ee
In these estimates, we considered (gapped) graphene $v \approx 10^6\,{\rm m/s}$ as model system and used a typical value of $ \kappa = 4\,{\rm eV}/\mu{\rm m^2} $, see estimates below. 

A suitable diagnostics of nonreciprocity, allowing direct access to the quantum dot resonances, is the local density of states $D(\varepsilon)$ inside the quantum dot. Naturally, $D(\varepsilon)$ can be obtained experimentally via the $dI/dV$ in STS measurements as in Refs.\cite{zhaoscience,crommie}. The quantity $D(\varepsilon)$ at $r = r_0$ can be conveniently written as the sum of $m$-state contributions $D(\varepsilon) = \sum_m D_m(\varepsilon)$, with 
\be
D_m (\varepsilon)=\sum_{\alpha} \langle |u_{\alpha}(r = r_0)|^2\rangle_{\lambda_d} \delta ( \varepsilon - \varepsilon_\alpha).
\label{eq:ddos}
\ee
Here $\alpha$ labels the radial eigenstates of Eq.(\ref{eq:Hnd}) for fixed $m$, and $\langle |u_\alpha(r = r_0)|^2 \rangle_{\lambda_d} = \int_0^\infty dr' |u_\alpha (r')|^2 e^{-(r'-r_0)^2/2 \lambda_d^2}$ represents a spatial average of the wavefunction centered at $r = r_0$. A gaussian weight is included in the density of states to account for the finite size of the tunneling region in real STS measurements\cite{zhaoscience,crommie}. 

\begin{figure}
\centering \includegraphics[scale=1.0]{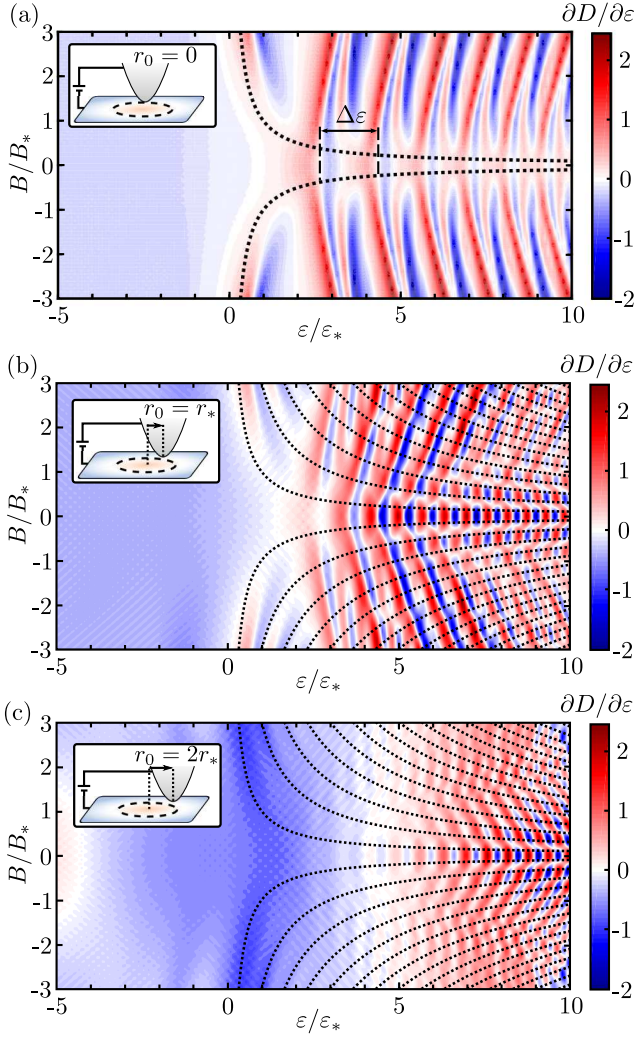}
\vspace{-3mm}
\caption{Maps of the local density of states as a function of position $r_0$ for a gapless Dirac quantum dot displaying   splitting of resonances in weak magnetic fields: (a) $r_0 = 0$, (b) $r_0 = r_*$, and (c) $r_0 = 2r_*$. Indicated with dotted lines is Eq.(\ref{eq:bstar}) for half-integer $m$. The off-centered spectral maps (b)-(c) are qualitatively different from the centered case (a) which is sensitive primarily to $m = \pm 1/2$ states. Characteristic units for magnetic field, $B_*$, is defined in Eq.(\ref{eq:units}). Plotted with dotted lines is Eq.(\ref{eq:bstar}) for half-integer $m$. To enhance spectral features, we plot in both panels the derivative of the local Density of States in Eq.(\ref{eq:ddos}).}
\label{fig:simulation}
\vspace{-5mm}
\end{figure}

\subsection{Splitting of quantum dot resonances}

Figure \ref{fig:simulation}(a) shows the resulting quantum dot spectrum as a function of $B$ for gapless Dirac systems, exhibiting the $B$-induced splitting of quantum dot resonances. In our calculations, we used $r_0 = 0$, $\lambda_d / r_* = 0.1$, and plotted $\partial D / \partial \varepsilon$ in Eq.(\ref{eq:ddos}) in order to enhance spectral features (see Appendix A for details). In agreement with our semiclassical interpretation, a half-period splitting is observed in the gapless spectral maps in Fig.\ref{fig:simulation}(a). Because in Fig.\ref{fig:simulation}(a) the wavefunction is probed at the center of the quantum dot, only small $m$ states ($m = \pm 1/2$) contribute to the spectral maps. It is important to stress that large $m$ states, which can be probed in off-centered STS measurement, are equally susceptible to the Berry phase splitting. Figure \ref{fig:simulation}(b) and (c) show such spectral maps, in which the wavefunctions are probed at (b) $r_0 = r_*$ and (c) $r_0 = 2 r_*$. In these cases, there is an overlap of peak splitting at different values of $B$, highlighted with fans of $B_{\rm c}$ in Eq.(\ref{eq:bstar}) for varying $m$ (dotted lines). At a larger value of $r_0$, states with larger $m$ and $\varepsilon$ can be probed. This is indicated by a larger contrast in the local density of states induced by such states in Fig.\ref{fig:simulation}(b) and (c). 

\begin{figure}
\centering \includegraphics[scale=1.0]{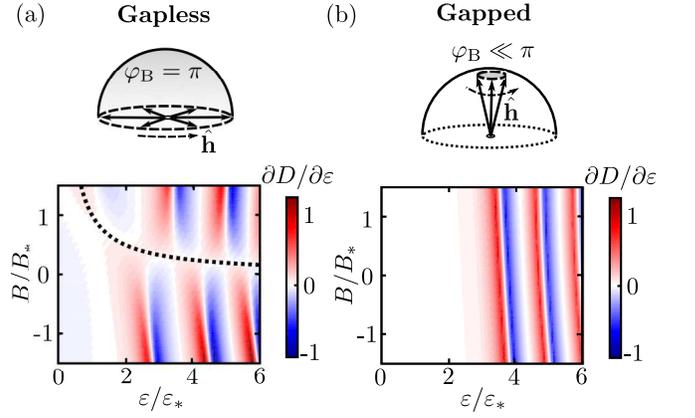}
\vspace{-7mm}
\caption{Partial-$m$ contribution to the on-center density of states for quantum dots in a) gapless and b) gapped Dirac systems.  The strong nonreciprocal effect induced by Berry phase disappears when a large gap $\Delta$ is opened. As a result, resonance splitting is dominated by (a) the Berry phase jump in gapless systems, and (b) orbital effects in gapped systems. The distinct behavior between (a) and (b) is shown in the partial $m=1/2$ maps of the density of states [indicated with a dotted line is Eq.(\ref{eq:bstar}) with $m=1/2$; $\Delta/\varepsilon_* = 5$ was used in (b)].}
\label{fig:gapeffect}
\vspace{-5mm}
\end{figure}

As shown in Fig.\ref{fig:gapeffect}, the splitting of the resonances for gapped systems is less prominent; in particular, splitting is dominated by the orbital contribution. Indeed, the peak splitting for the low-energy resonances in gapped Dirac systems ($\varepsilon\gtrsim\Delta$) can be quantified using a simple nonrelativistic model that is valid in the limit $\Delta \gg \varepsilon_*$. In this case, expansion of the Dirac Equation in powers of $\Delta$ gives a nonrelativistic Schr{\"o}dinger Equation for the first spinor component $\Psi_1({\bm r})$:
\be
\begin{array}{c}
\displaystyle \left[ {\bm q}^2/2\Delta + U({\bm r}) + \Delta  -eB/2\Delta \right] \Psi_1 = \varepsilon_{n,m}\Psi_1, \\ \\
\displaystyle \varepsilon_{n,m} = \hbar\omega  \left(2n + |m_-| +1 \right) - \mu_*  m_+ B.
\end{array}
\label{eq:spinless}
\ee
Here $\varepsilon_{n,m}$ are the quantized eigenvalues, $\omega = \sqrt{2\kappa/\Delta + e^2B^2 / 4\Delta^2}$, and $m_\pm = m \pm 1/2$. Here we introduced the orbital magnetic moment $\mu_* = e \hbar v^2 / 2 \Delta$, which can be viewed as an effective Bohr magneton of a free Dirc particle of mass $\Delta / v^2$. This term is responsible for the shift of the resonances with $B$ seen in Fig.\ref{fig:gapeffect}(b).

\subsection{Self-consistent calculation of the potential profile}

\begin{figure}
\centering \includegraphics[scale=1.0]{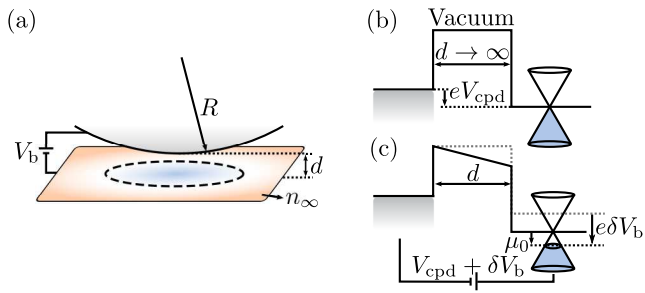}
\caption{(a) Schematics of the electrostatic model, showing a metallic sphere of radius $R$ separated a distance $d$ from the graphene plane. A potential bias $V_{\rm b}$ applied on the sphere induces a local variation of the carrier density, different from the carrier concentration density $n_\infty$ far from the sphere. (b,c) Band structure schematics showing band alignment between the metallic sphere and graphene for (b) large separation and (c) close proximity. Here $V_{\rm cpd}$ is the contact potential difference between graphene and the metallic sphere, $\delta V_{\rm b} = V_{\rm b} - V_{\rm cpd}$, and $\mu_ 0$ is the Fermi energy under the sphere.}
\label{fig:electrostatics}
\end{figure}

Estimates for $\kappa$ used in Eq.(\ref{eq:H}) can be obtained from a simple electrostatic model which involves a metallic sphere proximal to the graphene plane [Fig.\ref{fig:electrostatics}(a)]. This model accounts for the fields and charges induced by an STM tip on top of graphene, as discussed in Ref.[\onlinecite{zhaoscience}]. We denote with $R$ the metallic sphere radius, and with $d$ the sphere-graphene separation. A potential bias $\delta V_{\rm b}$ between the sphere and graphene [see Fig.\ref{fig:electrostatics}(b)--(c)] results in a spatially varying charge density profile 
\be
\delta n (r) \approx -\frac{e \delta V_{\rm b}+\mu(r)}{4\pi e^2 (d + r^2 / 2R)}.
\label{eq:electrostatics}
\ee
Here $\delta n(r) = {\rm sgn} [\mu (r)] \mu(r)^2 / \pi (\hbar v)^2 - n_{\infty}$ is the STM tip-induced charge density variation on graphene, with $\mu (r)$ the Fermi energy and $n_{\infty}$ the gate-induced carrier density far from the tip. Equation (\ref{eq:electrostatics}) is obtained from a parallel-plate capacitor model with a slowly varying plate separation $d_{\rm c}(r) \approx d + r^2 / 2R$. Higher-order terms arising due to the curvature of the electric field lines are neglected.

A straight-forward calculation yields a value of $\kappa = - \mu''(0)/2$ given by
\be
\kappa = - \frac{e\delta V_{\rm b}+\mu_0}{2Rd\sqrt{1+|\beta|}}.
\ee
Here $\mu_0$ is the Fermi energy directly under the sphere, and $\beta$ is a dimensionless number defined as
\be
\begin{array}{l}
\displaystyle \mu_0 = \frac{(\hbar v)^2}{8 e^2 d }\frac{1 - \sqrt{1+|\beta|}}{{\rm sgn}(\beta)}, \\
\displaystyle\beta = \frac{16e^2d}{(\hbar v)^2}\left[ e\delta V_{\rm b} -4\pi e^2 d n_{\infty}\right].
\end{array}
\ee
For typical values of $ R \sim 1\,\mu{\rm m}$, $d \sim 5\,{\rm nm}$, $\delta V_{\rm b} \sim 0.1\,{\rm V}$ and $n_{\infty} \sim 10^{11}\,{\rm cm^{-2}}$, we obtain the value of $\kappa \sim 4\cdot 10^{-6}\,{\rm eV/nm^2}$. 

\section{IV. Discussion}

We point out that the nonreciprocity mechanism considered above is inherent to Dirac materials. In contrast, Faraday and Kerr rotation, two notable examples of nonreciprocity which can be sizable in two dimensional materials such as graphene\cite{faraday1,faraday2}, are also present in general semiconductor materials. The same applies to magnetoplasmonic effects, e.g. unidirectional low frequency edge modes\cite{magnetoplasmontheory,magnetoplasmontheory2,plasmon1,plasmon2,plasmon3}, which are also present in generic two-dimensional structures\cite{magnetoplasmontheory}. 

The large magnitude and tunability of our nonreciprocal effect may help design optical devices, such as nanoscale isolators and circulators, which are driven by Berry phase. Of special interest are photonic effects in Dirac quantum dots. Indeed, electrostatic doping can, via the Pauli blocking mechanism, induce a strong and tunable electron-photon coupling. This, combined with the in situ tunability of the resonance dispersion\cite{zhaoscience}, can make Dirac quantum dots useful in miniaturizing nanophotonic systems. 

It is instructive to compare our nonreciprocal effect with other exotic manifestations of Berry phase predicted to occur in Dirac systems, such as Berry phase modification to exciton spectra \cite{exciton1,exciton2}, optical gyrotropy induced by Berry's phase \cite{gyrotropy} and chiral plasmon in gapped Dirac systems \cite{chiralplasmons1,chiralplasmons2}. In realistic electronic systems electron decoherence usually hinders observation of such subtle effects. We therefore expect that the quantum dot states readily available in Dirac materials provide a new and optimal setting for locally probing Berry phase physics. 

Since our predictions, such as the strong dependence of resonance splitting on $\Delta$, only rely on the Dirac nature of charge carriers, they can be tested in a wide range of materials. Graphene is the prototypical material to explore the case $\Delta = 0 $; graphene on top of axis-aligned hBN substrate allows to explore the case $\Delta \sim 50\,{\rm meV}$ \cite{hbngap1,hbngap2}; monolayers of transition metal dichalcogenides such as MoS$_2$ allow to explore $\Delta$ of an eV scale\cite{tmdexciton,tmdsreview,heinzreview}. Furthermore, the value of $\varepsilon_*$ can also be tuned with electrostatic potential, as demonstrated in Ref.\cite{zhaoscience}.

\section{Summary}

To summarize, quantum dots embedded in Dirac materials grant access to a new nonreciprocity mechanism originating from the Berry phase. This mechanism, which is unique to Dirac materials, leads to stronger nonreciprocity than that for other known mechanisms. The anomalous strength of the effect and its in situ tunability makes Dirac quantum dots an appealing platform for nonreciprocal nanophotonics. The recent introduction of Dirac quantum dots in graphene makes these predictions easily testable in on-going experiments.

\section{Acknowledgements}

We thank M. S. Dresselhaus, J. Stroscio and B. Skinner for advice and valuable discussions, and acknowledge support by the National Science Foundation Grant No. DMR-1004147 [J.F.R.-N.] and by the Center for Excitonics, an Energy Frontier Research Center funded by the US Department of Energy, Office of Science, Basic Energy Sciences under Award No. DE-SC0001088 [LL]. 

\subsection{Appendix A: Computational details}

To solve Eq.(\ref{eq:Hnd}) above, we use the finite difference method in the interval $0 < r < L$. The azimuthal quantum numbers are chosen in a finite range, $-M \le m \le M$, with $M$ large enough to represent accurately the states in the energy range of interest. In our calculations, we used a system of size $L/r_* = 10$ discretized in $N = 600$ lattice sites, with maximum azimuthal quantum number $M = 31/2$. To calculate the density of states, Eq.(\ref{eq:ddos}), we approximate the delta-function $\delta (\varepsilon)$ by a Lorentzian $\delta (\varepsilon) \approx \Gamma / \pi(\varepsilon^2+\Gamma^2)$. We used a broadening parameter $\Gamma / \varepsilon_* = 0.25$, and set a Gaussian weight in the spatial average $\langle\ldots\rangle_{\lambda_d}$ of the wavefunction to $\lambda_d / r_* = 0.1$.


\begin{thebibliography}{99}

\bibitem{zhaoscience} Y. Zhao, J. Wyrick, F. D. Natterer, J. F. Rodriguez-Nieva, C. Lewandowski, K. Watanabe, T. Taniguchi, L. S. Levitov, N. B. Zhitenev, and J. A. Stroscio, Science {\bf 348}, 672 (2015).

\bibitem{crommie} J. Lee, D. Wong, J. Velasco Jr., J.~F. Rodriguez-Nieva, S. Kahn, H.-Z. Tsai, T. Taniguchi, K. Watanabe, A. Zettl, F. Wang, L.~S. Levitov, and M.~F. Crommie, Nat. Phys. {\bf 12}, 1032 (2016).

\bibitem{fockstates} H.-Y. Chen, V. Apalkov, and T. Chakraborty, Phys. Rev. Lett. {\bf 98}, 186803 (2007).

\bibitem{causticsqd} J. Cserti, A. P{\'a}lyi, and C. P{\'e}terfalvi, Phys. Rev. Lett. {\bf 99}, 246801 (2007).

\bibitem{integrableqd} J. H. Bardarson, M. Titov, and P. W. Brouwer, Phys. Rev. Lett. {\bf 102}, 226803 (2009). 

\bibitem{quantumdot} P. G. Silvestrov and K. B. Efetov, Phys. Rev. Lett. {\bf 98}, 016802 (2007).

\bibitem{qdboundstates} A. Matulis, and F. M. Peeters, Phys. Rev. B {\bf 77}, 115423 (2008). 

\bibitem{scatteringqd} J.-S. Wu, and M. M. Fogler, Phys. Rev. B {\bf 90}, 235402 (2014). 

\bibitem{klein1} M. I. Katsnelson, K. S. Novoselov, and A. K. Geim, Nat. Phys. {\bf 2}, 620 (2006).

\bibitem{klein2} V. V. Cheianov and V. I. Fal'ko, Phys. Rev. B. {\bf 74}, 041403 (2006).

\bibitem{berry} M. V. Berry, Proc. R. Soc. Lond. A {\bf 392}, 45 (1984). 

\bibitem{niu2} G. Sundaram and Q. Niu, Phys. Rev. B {\bf 59}, 14915 (1999). 

\bibitem{graphene4} A. V. Shytov, M. S. Rudner, and L. S. Levitov, Phys. Rev. Lett. {\bf 101}, 156804 (2008). 


\bibitem{changreview} M.-C. Chang and Q. Niu, J. Phys: Condens. Matter {\bf 20}, 193202 (2008).

\bibitem{graphene3} A. F. Young and P. Kim, Nat. Phys. {\bf 5}, 222 (2009). 


\bibitem{niu} D. Xiao, M.-C. Chang, Q. Niu, Rev. Mod. Phys. {\bf 82}, 1959 (2010).

\bibitem{changbp} M.-C. Chang and Q. Niu, Phys. Rev. B {\bf 53}, 7010 (1996).

\bibitem{fan} Z. Yu and S. Fan, Nat. Photon. {\bf 3}, 91 (2009).
  
\bibitem{ross} L. Bi, J. Hu, P. Jiang, D. H. Kim, G. F. Dionne, L. C. Kimerling, and C. A. Ross, Nat. Photon. {\bf 5}, 758 (2011).
 
\bibitem{stone} A. D. Stone, Physics Today {\bf 58}, 37 (2005).

\bibitem{graphene1} K. S. Novoselov, A. K. Geim, S.V. Morozov, D. Jiang, M. I. Katsnelson, I. V. Grigorieva, S. V. Dubonos, and A. A. Firsov, Nature {\bf 438}, 197 (2005). 

\bibitem{graphene2} Y. Zhang, Y.-W. Tan, H. L. Stormer, and P. Kim, Nature {\bf 438}, 201 (2005). 

\bibitem{bs} This method of quantization on the surface of an invariant tori is known as Einstein-Brillouin-Keller (EBK) quantization rule. It differs from the usual way of quantizing using closed orbits obtained from semiclassical equations of motion. The EBK method applies to completely integrable $d$-dimensional systems as well as chaotic systems. For more details, see Ref.\cite{stone}. 

\bibitem{faraday1} I. Crassee, J. Levallois, A. L. Walter, M. Ostler, A. Bostwick, E. Rotenberg, T. Seyller, D. van der Marel, and A. B. Kuzmenko, Nat. Phys. {\bf 7}, 48 (2011).

\bibitem{faraday2} R. Shimano, G. Yumoto, J. Y. Yoo, R. Matsunaga, S. Tanabe, H. Hibino, T. Morimoto, and H. Aoki, Nat. Commun. {\bf 4}, 1841 (2013).

\bibitem{magnetoplasmontheory} V. A. Volkov, and S. A. Mikhailov, Sov. Phys. JETP {\bf 67}, 1639 (1988). 

\bibitem{magnetoplasmontheory2} W. Wang, S. P. Apell, J. M. Kinaret, Phys. Rev. B {\bf 86}, 125450 (2012). 

\bibitem{plasmon1} I. Crassee, M. Orlita, M. Potemski, A. L. Walter, M. Ostler, Th. Seyller, I. Gaponenko, J. Chen, and A. B. Kuzmenko, Nano Lett. {\bf 12}, 2470 (2012).

\bibitem{plasmon2} H. Yan, Z. Li, X. Li, W. Zhu, P. Avouris, and F. Xia, Nano Lett. {\bf 12}, 3766 (2012).

\bibitem{plasmon3} I. Petkovi{\'c}, F. I. B. Williams, K. Bennaceur, F. Portier, P. Roche, and D. C. Glattli, Phys. Rev. Lett. {\bf 110}, 016801 (2013). 

\bibitem{exciton1} A. Srivastava and A. Imamo\u{g}lu, Phys. Rev. Lett. {\bf 115}, 166802 (2015). 

\bibitem{exciton2} J. Zhou, W.-Y. Shan, W. Yao, and D. Xiao, Phys. Rev. Lett. {\bf 115}, 166803 (2015).

\bibitem{gyrotropy} S. Zhong, J. Orenstein, and J.~E. Moore, Phys. Rev. Lett. {\bf 115}, 117403 (2015).

\bibitem{chiralplasmons1} A. Kumar, A. Nemilentsau, K.~H. Fung, G. Hanson, N.~X. Fang, and T. Low, Phys. Rev. B {\bf 93}, 041413 (2016).

\bibitem{chiralplasmons2} J.~C.~W. Song and M.~S. Rudner, PNAS {\bf 113}, 4658 (2016). 

\bibitem{hbngap1} B. Hunt, J. D. Sanchez-Yamagishi, A. F. Young, M. Yankowitz, B. J. LeRoy, K. Watanabe, T. Taniguchi, P. Moon, M. Koshino, P. Jarillo-Herrero, and R. C. Ashoori, Science {\bf 340}, 1427 (2013).

\bibitem{hbngap2} C. R. Woods, L. Britnell, A. Eckmann, R. S. Ma, J. C. Lu, H. M. Guo, X. Lin, G. L. Yu, Y. Cao, R. V. Gorbachev, A. V. Kretinin, J. Park, L. A. Ponomarenko, M. I. Katsnelson, Y. N. Gornostyrev, K. Watanabe, T. Taniguchi, C. Casiraghi, H.-J. Gao, A. K. Geim, K. S. Novoselov, Nat. Phys. {\bf 10}, 451 (2014).

\bibitem{tmdexciton} A. Ramasubramaniam, Phys. Rev. B {\bf 86}, 115409 (2012). 

\bibitem{tmdsreview} Q. H. Wang, K. Kalantar-Zadeh, A. Kis, J. N. Coleman, and M. S. Strano, Nat. Nanotech {\bf 7}, 699 (2012). 

\bibitem{heinzreview} X. Xu, W. Yao, D. Xiao, and T. F. Heinz, Nat. Phys. {\bf 10}, 343 (2014). 
  
\end{thebibliography}
\end{document}